\documentclass[prb,aps,twocolumn,superscriptaddress,showpacs]{revtex4}
\usepackage{graphicx, amsmath, amsthm}
\makeatletter
\begin{document}
\title{Impurity-induced configuration-transition in the Fulde-Ferrell-Larkin-Ovchinnikov state of a $d$-wave superconductor}

\author{Qian Wang}
\affiliation{Texas Center for Superconductivity, University of Houston,
Houston, Texas 77204}
\author{Chia-Ren Hu}
\affiliation{Department of Physics, Texas A\&M University, College
Station,
Texas 77843}
\author{Chin-Sen Ting}
\affiliation{Texas Center for Superconductivity, University of Houston,
Houston, Texas 77204}
\date{\today}
\begin{abstract}
The effect of impurities on Fulde-Ferrell-Larkin-Ovchinnikov (FFLO) states in a layered superconductor with $d$-wave pairing symmetry is investigated using the tight-binding model and the Bogoliubov-de Gennes equations. 
At low temperature and a strong exchange or Zeeman field applied parallel to its conducting plane, a two-dimensional (2D) square lattice-like Larkin-Ovchinnikov state is more energetically favorable in a clean system. In the presence of impurities, the spatial profile of the order parameter remains as a 2D square lattice, and it is distorted only near the impurities when the impurity concentration is low.  As impurity concentration is increased to a certain level, quasi-one-dimensional (1D) like FFLO state becomes more energetically favorable. Increasing temperature with fixed impurity concentration can also induce a 2D to 1D FFLO state transition. Within the present finite size calculation, we did not find the existence of the Fulde and Ferrell state before the system becoming normal as the impurity concentration is raised. 
   
\end{abstract}
\maketitle

Recently, the possible existence of the  Fulde-Ferrell-Larkin-Ovchinnikov (FFLO) state in heavy fermion compound CeCoIn$_5$~\cite{FFLO-CeCoIn5} has revived the interest in this inhomogeneous superconducting state. The FFLO state was predicted forty years ago by Fulde and Ferrell~\cite{FF}, and Larkin and Ovchinnikov~\cite{LO}.  for a superconductor subject to a strong exchange field or a strong magnetic (Zeeman) field applied in the planes of a quasi-two-dimensional superconductor. Soon after the prediction, it was shown that the existence of the FFLO state depends on the magnitude of the exchange or Zeeman field and is very sensitive to the presence of impurities~\cite{Aslamazov}. This could be one of the reasons that the FFLO state has never been observed in conventional low-temperature superconductors.

CeCoIn$_5$ is likely to be a quasi-two-dimensional $d$-wave superconductor~\cite{dSC-CeCoIn5}. It has been predicted that at low temperature in a clean two dimensional $d$-wave superconductor, the order parameter of the energetically favored Larkin-Ovchinnikov (LO) state has 2D square-lattice-like structure~\cite{Maki, Shimahara, qwang05}, while the 1D stripe-like LO state is more favored in $s$-wave superconductors. In the LO state, the order parameter is real and spatially inhomogeneous with positive and negative signs spreading over the configuration space. Whereas in the FF state, the order parameter gets a spatially inhomogeneous phase similar to that of  a plane-wave, but its magnitude remains to be spatially uniform. In such a state, a spin current has been shown to exist. In the clean limit, CeCoIn$_5$ is a material with very long mean free path. However, it is possible to introduce impurities in this material by substituting some Ce atoms by La atoms~\cite{Petrovic02}.  The scattering from impurities may break the symmetry of the order parameter in the FFLO state. Using Born approximation to treat impurity scattering, Agterberg and Yang  have derived a Ginzburg-Landau (GL) free energy functional for both $s$-wave and $d$-wave superconductors with non-magnetic impurities and Zeeman fields~\cite{Yang}. Using this free energy, they have shown that impurities could induce a change in the structure of the FFLO state in a $d$-wave superconductor, namely, a phase transition from an LO state in the clean limit to a FF state with relatively high concentration of impurities.  This GL free energy is valid near normal to superconductor phase transition line in the (T, H) plane. The impurity induced FF phase was found to be between the LO phase and the uniform superconducting phase and thus is away from the normal to superconductor phase transition line. Therefore, it is of interest to study such a problem with an approach which goes beyond the GL theory, and to take the effect due to multiple scatterings from impurities into account when the impurity concentration is relatively high
.

In this paper, based upon a tight-binding model for a $d$-wave superconductor with FFLO state, we show that when impurity concentration is increased to a certain level, the spatial profile of the order parameter changes from 2D like to 1D like.  This 1D LO phase lies between 2D FFLO (or LO) phase and the normal phase. No FF state before the system becoming the normal state is found.   In the 1D LO phase, the impurities would like to reside on or very close to the nodal lines in real space. As a result, the nodal lines would not be as straight as those in the clean limit.    When temperature is raised and with fixed impurity concentration, the mean-free path should decrease and a 2D to 1D LO state phase transition is shown to occur as well.       

To investigate the effect of impurities on the FFLO state in superconductors, we study a phenomenological model on a square lattice: 
\begin{equation}
\begin{split}
H&=\sum_{<i,j>,\sigma}-t_{ij} c^\dagger_{i\sigma}c_{j\sigma}-\sum_{i,\sigma}(\mu+\sigma h+U_0\delta_{i,j_m})c^\dagger_{i\sigma}c_{i\sigma}\\
&+\sum_{i,j}(\Delta_{ij}c^\dagger_{i\sigma}c_{j\sigma}^\dagger+H.c.),
\end{split}
\end{equation}
where $t_{ij}$ is the hopping integral and $\mu$ is the chemical potential. Throughout our study, we take $t_{ij}$ to be unity for nearest neighbors and zero otherwise. The Zeeman energy term $\sigma h$, with $\sigma=\pm1$ for spin-up and -down electrons, arises from the interaction between the magnetic field and the spins of electrons. $U_0$ denotes the impurity potential. $j_m$, with $m=1$ to $n$, where $n$ is the number of impurities. The singlet superconducting order parameter has the following definition: $\Delta_{ij}=V<c_{i\uparrow}c_{j\downarrow}-c_{i\downarrow}c_{j\uparrow}>/2$.  

This Hamiltonian can be diagonalized by solving the discrete Bogoliubov-de Gennes equations:
\begin{equation}
\sum_j\left({{\cal H}_{ij,\sigma}\atop \Delta_{ij}^*}
        {\Delta_{ij}\atop{-{\cal H}_{ij,\bar\sigma}^*}}\right)
        \left(u_{j\sigma}^n\atop v_{j\bar\sigma}^n\right) =
        E_n\left(u_{j\sigma}^n\atop v_{j\bar\sigma}^n \right),
\label{discreteBdG}
\end{equation}
where
$H_{ij,\sigma} = -t_{ij} - (\mu+\sigma h)\delta_{ij}+U_0\delta_{i,j_m}$. $u_{j\sigma}^n$ and $v_{j\bar\sigma}^{n}$ are the Bogoliubov
quasiparticle amplitudes on the $j$-th site. The self-consistency
condition for the OP:
\begin{equation}
\Delta_{ij} =
\delta_{j,i+\gamma}\frac{V}{4}\sum_n\tanh\frac{E_n}{2k_BT}(u^n_{i\uparrow}
    v^{n*}_{j\downarrow}+u^n_{j\downarrow}v^{n*}_{i\uparrow})\,.
\label{selfconsist}
\end{equation}
is solved by iteration. Here $\gamma = (\pm 1,0)$ and $(0,\pm 1)$, and
$\Delta_i=(\Delta_{i+\hat x} +
\Delta_{i-\hat x} - \Delta_{i+\hat y} - \Delta_{i-\hat y})/4$
is the d-SC OP at site $i$.

 We choose $U_0=5.0$ for a relatively strong impurity potential. If the impurity potential is taken to be too high, the LO state may be destroyed before we can see a 2D to 1D transition. The size of the lattice for performing the numerical computation is chosen to be $32\times32$. There are totally $N=1024$ sites. The impurities are randomly distributed. For each impurity concentration, we calculate the order parameter for several independent set of impurity configurations. The LO phase transition point for a fixed impurity concentration should be obtained by taking the average of all trial transition points in different impurity configurations. In order to check whether the FF state becomes the ground state when impurity concentration is increased, we assign a random phase to the order parameter at each site as an initial condition. 

Impurities suppress the superconductivity in a $d$-wave superconductor and they may also change the periodicity of the order parameter in the LO state. We take $V=2.0$, $\mu=-1.0$ and $h=0.40$ in all the calculations. With this set of parameter, the order parameter shows 2D square-lattice variation for a clean $d$-wave superconductor at  low temperature. In our numerical calculation with impurities, we assume that the periodicity of the LO state along the  $x$ direction could be different from that of the $y$ direction.  

\begin{figure}
\centering
\vskip 3.8in
\includegraphics[width=2.5in]{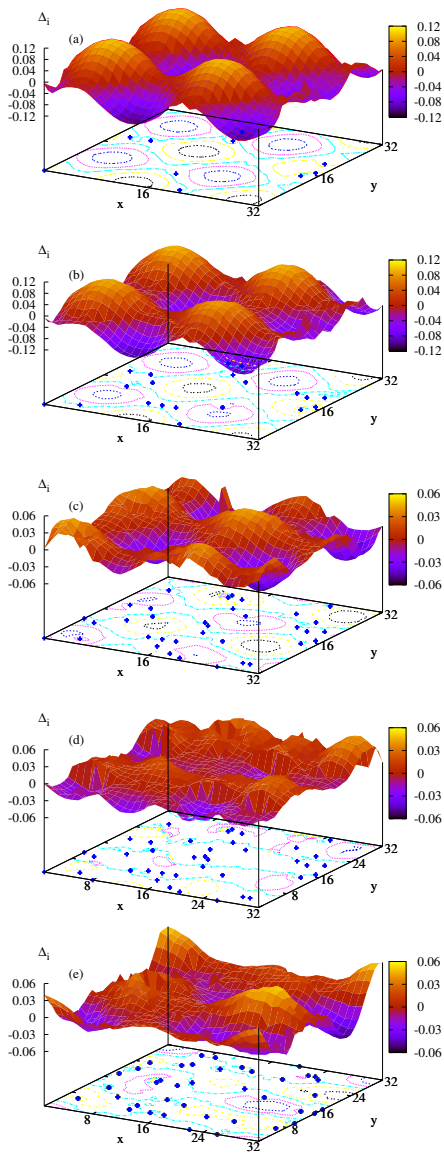}
\vskip 0.2in
\caption{Spatial profiles of the order parameter for various impurity concentrations. The size of the lattice is 32$\times$32.  (a) 10 impurities, (b) 20 impurities, (c) 30 impurities, (d) 40 impurities, and (e) 40 impurities. Here (d) and (e) have different impurity configurations. When there are more than 35 impurities, the nodal lines no longer cross each other. The sites of impurities are marked by ''+'' on the contour plots.}
\label{fig:im}
\end{figure}

{\it Configuration space phase transition induced by impurities in the low temperature limit:}
Fig.~\ref{fig:im} shows the calculated order parameter for various impurity concentrations. 
When the concentration of impurities is very low (below $n/N\approx 3.0\%$), the order parameter still iterates practically
to that of a 2D lattice LO state  as that in a clean limit with nodal lines along (110) directions, except the distortion occurs near the impurity site, as shown in Fig.~\ref{fig:im} (a) and (b) for $n/N\approx 1.0\%$ and  $2.0\%$. Impurities are found likely to stay on or near nodal lines of the order parameter in the LO state. Looking at the spatial profiles of the order parameter, one can hardly find the impurity sites in these graphs because they stay at places where the order parameter is very small or zero. As the impurity concentration increases, the magnitude of the order parameter decreases and at the same time, the distortions become stronger and stronger. When the concentration reaches a certain level (above 3.5\% in our case) as shown in Fig.~\ref{fig:im} (d) and (e) for $n/N\approx 4.0\%$, the order parameter becomes quasi-one-dimensional.  The nodal lines no longer cross each other and are approximately along (100) direction. Further increasing of impurity concentration leads to the reduction of the superconducting order parameter until it vanishes. In the above graphs, the impurity site is marked by a "+" in the corresponding contour plots. The configurations for the positions of impurities are generated randomly by computers. We also have performed the same calculation for 5 other configurations with the same number of impurities. The essential feature of the above conclusion remains practically unchanged. For the purpose of illustration, we present the results for two different impurity configurations with 40 impurities in our lattice as shown in  Fig.~\ref{fig:im} (d) and 1(e). However, unlike the results in Ref.~10, we did not obtain an FF phase in our calculation. The phases of the order parameter at different sites are always converging to the same value in the self-consistent calculation. We have also investigated this effect for an s-wave superconductor. In a clean system, the energetically favored state is 1D stripe-like LO state. Upon increasing impurity concentration, the order parameter remains 1D stripe-like until it vanishes and becomes normal state.   An FF phase has never shown up in the present approach.

\begin{figure}
\centering
\vskip 2.5in
\includegraphics[width=2.5in] {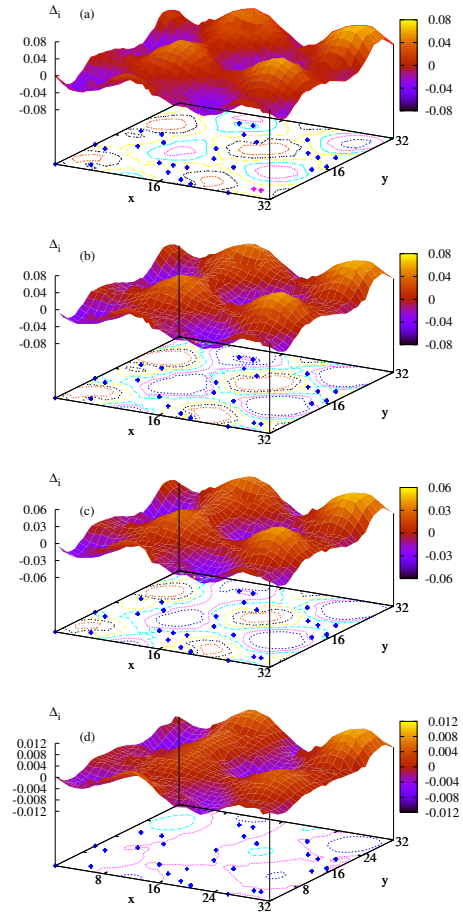}
\caption{ Spatial variations of $\Delta_i$ at various temperature with 30 impurities on a lattice of size $32\times 32$:(a) T=0.001; (b) T=0.005; (c) T=0.01; (d) T=0.02. The impurities are marked by ''+'' on the contour plots.}
\label{fig:im1}
\end{figure}

{\it Configuration space phase transition in the presence of impurities due to the raise of temperatures:}  To study this effect, we fix the number of impurities in the lattice to 30. The results are shown in Fig.~\ref{fig:im1}. At low temperature (T=0.001), the order parameter is clearly having a 2D structure even though it is strongly distorted at or near the sites of impurities. As the temperature is raised, we see the variation becomes more and more 1D like. At T=0.021, the LO state is destroyed, and the system becomes normal. For lower impurity concentration, say $~2.0\%$, we found no 2D to 1D transition. The 2D LO state persists up close to the superconducting transition temperature where the order parameter vanishes.

The phase change caused by raising the temperature indicates that in a T-H phase diagram, the 1D LO phase locates between the 2D LO phase and the normal state, while the FF state predicted in Ref.~10 lies between the LO state and the uniform superconducting state. The two studies investigate two different regions in the phase diagram. Our result is in the high field region resulting from the fact that our method is more suitable for small period FFLO states in strong exchange or Zeeman field, while the study in Ref.~10 might be in relatively low field.  For low field, the periodicity of the LO state will be large, the required lattice size is beyond our numerical capability.

{\it The Local density of states for spin up quasiparticles (LDOS) at a site next to an impurity when is impurity concentration low:}
With impurity concentration is increased to $3.5\%$, our result shows a 2D LO state to 1D LO state transition. We would like to check whether this transition could be observed by the scanning tunneling microscopy experiments, which directly measure the LDOS of the system and its spatial maps. 
In Ref. 8, the present authors have calculated the LDOS at a site nearest neighboring to a single impurity when the impurity is located at an saddle point or extreme of the order parameter. The LDOS spectrum at site nearest neighboring to the impurity show quite different behavior from a pure $d$-wave superconductor. If the impurity is at an order parameter saddle
point,  the LDOS is practically unaffected by the impurity, with no impurity-induced resonant peak or peaks appearing.  This is because the order parameter seen by a quasiparticle before and after scattering by an impurity does not change sign, thus there are no zero energy Andreev's bound states.  If the impurity is at an order parameter extremum, a $\pm\epsilon_0 +$ (Zeeman energy) pair of finite-energy resonance peaks are induced in the LDOS by the impurity, instead of one at near zero energy, as in a uniform $d$-wave superconductor. When multiple impurities are considered, the impurities can not all locate at a saddle point or a order parameter extremum. Unless a impurity is located at one of the  saddle points, we would expect that the LDOS spectrum would be similar to the case when the impurity is at an order parameter extreme because one can always find a quasiclassical orbit of incident and scattered quasiparticles experiencing sign change of the order parameter. 

\begin{figure}
\centering
\vskip 0.2in
\includegraphics[width=2.5in]{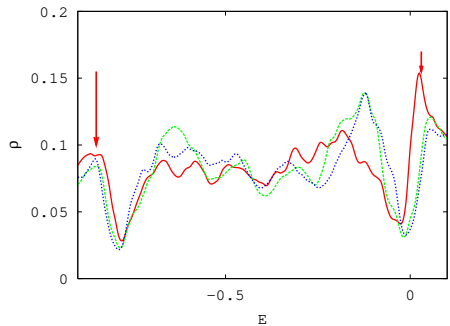}
\caption{the LDOS at sites next to three impurity sites when there are 10 impurities in a lattice of  size 32$\times$32. $V=2.0$, $h=0.4$. The outside two peaks for each curve, marked by red arrows, are coherence peaks. The subgap LDOS are complicated due to the interferences among the resonance states formed in the vicinity of different impurity sites.}
\label{fig:ldos}
\end{figure}

In fig.~\ref{fig:ldos}, we show the LDOS at sites nearest neighboring to some impurity sites when impurity concentration is low (10 impurities in a lattice of the size 32$\times$32 or the impurity concentration is approximately $1\%$). The subgap LDOS at these sites are very complicated comparing to the case of single impurity. This is due to the interferences among the resonance states formed in the vicinity of different impurities sites. Again the stable configuration here for the order parameter is still two-dimensional, and the image maps of the LDOS at certain fixed energies still exhibit the same checkerboard patterns~\cite{qwang06}. However, near the 2D to 1D transitional region, the impurity concentration becomes larger and it is between 3 to 4\%. Under this condition, there are strong interferences among the states from different impurities and nodal lines. As a result, no clear image maps could be obtained to show the transition. This conclusion also hold true if we fix the impurity concentration at $3\%$, then raise the temperature to detect the 2D to 1D transition. Since the magnetization along the nodal lines should be larger~\cite{qwang06}, the 2D to 1D transition could be observed by neutron scattering experiments. 


In summary, the effect of non-magnetic impurities on the FFLO state in a $d$-wave superconductor is studied by using the tight-binding model and the Bogoliubov-de Gennes equations. We have shown that the impurities can induce a 2D to 1D configuration space phase transition in such a system. 
At low temperature, with a strong magnetic field applied to its conducting plane, a 2D lattice like Larkin-Ovchinnikov state is more energetically favorable in a clean system. When impurity concentration is very low, the spatial profile of the order parameter remains practically unchanged, except it  is distorted near the impurity sites.  As impurity concentration is raised to a certain level, the scatterings from impurities break the symmetry of the system and 1D-stripe like LO state becomes more energetically favorable. The 1D-stripes are not straight because they have to accommodate the impurities to reside on or near the nodal lines. The 2D to 1D LO  transition is also obtained if the impurity concentration is fixed at, say $~3.0\%$ and the temperature is raised. In the present self-consistent calculations on square lattice with size 32x32, we are not able to find the FF state as predicted in Ref.~10.

This work is supported by a grant from the Robert A. Welch Foundation
under NO. E-1146 and by the Texas Center for Superconductivity at the
University of Houston through the State of Texas.

\end{document}